\documentclass[12pt]{article}
\usepackage{amsfonts,amsmath}
\usepackage{amssymb}

\makeatletter \@addtoreset{equation}{section} \makeatother

\newcommand{\dr}{{\rm d}}
{\vspace{3mm} }

\def\al{\alpha}

\def\*{\star}

\def\E2{\mathbf{E}}

\newcommand{\be}{\begin{equation}}
\newcommand{\ee}{\end{equation}}
\newcommand{\bee}{\begin{eqnarray}}
\newcommand{\beee}{\begin{array}}
\newcommand{\eee}{\end{eqnarray}}
\newcommand{\eeee}{\end{array}}
\newcommand{\ga}{\alpha}
\newcommand{\gb}{\beta}
\newcommand{\gga}{\gamma}

\newcommand{\D}{{\cal D}}

\newcommand{\W}{{\cal W}}
\newcommand{\B}{{\cal B}}

\newcommand{\gd}{\delta}

\newcommand{\gk}{\varkappa}
\newcommand{\gep}{\epsilon}

\newcommand{\go}{\omega}

\newcommand{\dal}{\dot \alpha}
\newcommand{\dgb}{\dot \beta}
\newcommand{\dgga}{\dot \gamma}

\newcommand{\p}{\partial}
\renewcommand{\L}{\cal L}

\newcommand{\ff}{\frac}

\begin{document}
\begin{flushright}
FIAN/TD/2017-29\\
\end{flushright}

\vspace{0.5cm}
\begin{center}
{\large\bf Lorentz covariant form of extended higher-spin equations}

\vspace{1 cm}

\textbf{V.E.~Didenko,}\textsc{$^{1}$}\textbf{
N.G.~Misuna}\textsc{$^{1,2}$}\textbf{
and M.A.~Vasiliev}\textsc{$^{1}$}\\
 \vspace{0.5cm}
 \textsc{$^{1}$} \textit{I.E. Tamm Department of Theoretical Physics,
Lebedev Physical Institute,}\\
 \textit{ Leninsky prospect 53, 119991, Moscow, Russia}\\

\par\end{center}

\begin{center}
\textsc{$^{2}$}\textit{Moscow Institute of Physics and Technology,}\\
 \textit{ Institutsky lane 9, 141700, Dolgoprudny, Moscow region,
Russia}\\

\par\end{center}

\begin{center}
\vspace{0.6cm}
 didenko@lpi.ru, misuna@phystech.edu, vasiliev@lpi.ru \\

\par\end{center}

\vspace{0.4cm}

\begin{abstract}
\noindent The extension of nonlinear higher-spin equations in
$d=4$  proposed in [arXiv:1504.07289] for the construction of
invariant functional is shown to respect local Lorentz symmetry.
The equations are rewritten in a manifestly Lorentz covariant form
resulting  from some Stueckelberg-like field transformation. We
also show that the two field-independent central terms entering
higher-spin equations which are not entirely fixed by the
consistency alone get fixed unambiguously by the requirement of
Lorentz symmetry. One of the important advantages of the proposed
approach demonstrated in the paper is the remarkable
simplification of the perturbative analysis.
\end{abstract}

\section{Introduction}
Nonlinear higher-spin (HS) equations were presented
\cite{Vasiliev:1990pl, more} in the unfolded form of covariant
first-order  differential  equations. Generalizing
Maurer-Cartan equations, unfolded equations \cite{Vasiliev:88,Vasiliev:89}
(for  reviews see e.g. \cite{Vasiliev:Rev,Boulanger:2008up})
manifestly control gauge symmetries and diffeomorphism
invariance. This is achieved due to  appearance of infinite towers of
auxiliary fields packed into HS master fields, that describe
all on-shell derivatives of HS fields. Schematically any unfolded system has the
form
\be\label{unf}
\dr W(x)=F(W(x))\,,
\ee
where $W(x)$ is some set of differential form fields and $\dr$ is
the space-time De Rham differential. (All products are wedge
products. The wedge symbol is implicit.) Unfolded equations
\eqref{unf} are consistent if no other than \eqref{unf} conditions
result from the integrability requirement $\dr^2=0$.

Among different approaches to HS dynamics, one of the most natural
is the metric-like one pioneered by Fronsdal \cite{Fronsdal}.
However, even though it aims at generalization of the Einstein
gravity in terms of habitual tensor fields, the progress in the
construction of interactions  is rather limited beyond the cubic
(Lagrangian) order (see e.g.
\cite{Berends:1984rq}-\cite{Francia:2016weg}) due to the technical
challenge and lack of geometric intuition within this formalism.
At least to lower orders, however, the HS $AdS/CFT$ holography
allows reconstructing local HS interactions
\cite{Sleight:2016dba}. The situation with the light-cone approach
is analogous \cite{Bengtsson:1983pd}-\cite{Ponomarev:2017ar} with
the notable exception for the paper by Metsaev
\cite{Metsaev:1991mt} who was able to deduce nontrivial
restrictions on the cubic vertex from the analysis of quartic
interactions.

The frame-like approach
initiated in \cite{Vasiliev:1980as,Vasiliev:1986td} is a HS
generalization of the Cartan formulation of gravity. It naturally
generalizes the spin-two curvature two-forms to higher spins the
same time bringing in a new piece of information such as the
notion of HS algebra and tightly related ingredient, spectrum of
auxiliary fields. In the frame-like approach both covariant HS
cubic couplings  \cite{Fradkin:1987ks} and fully nonlinear HS
equations were  found \cite{Vasiliev:1990pl, more}. Its relation
to the metric-like language is due to the local Lorentz symmetry
which enables one rewriting field components from the frame-like
setting to the metric-like and vice versa. (Recall that in Cartan's
gravity
 the equivalence principle rests on both diffeomorphism invariance and
local Lorentz symmetry.)

Retrospectively, in formulating nonlinear HS equations the guiding
rules were the consistency of \eqref{unf} and the correct free
limit about $AdS$ background which is the exact vacuum solution.
The former automatically guarantees gauge invariance.
Diffeomorphism invariance was by construction. Local Lorentz
symmetry was also inbuilt by considering the problem in terms of
multispinors as finite-dimensional Lorentz modules. All these
properties
 were inherited by the nonlinear HS equations of \cite{Vasiliev:1990pl, more}.
Somewhat later it was realized \cite{Vasiliev:Rev}
that the form of equations of \cite{more} is to large
extent fixed by the local Lorentz covariance principle that can be easily lost
by an otherwise consistent deformation of the equations of \cite{more}.

 Local  Lorentz covariance  implies that the equations should be of the  form
\be\label{lorunf}
D^{L}W=F^{L}(W)\,,
\ee
 where $D^{L}=\dr+\go^{L}$ is the Lorentz
covariant derivative acting in the standard way on any Lorentz
tensor (multispinor) with the convention that $D^{L}\go^L$ is identified
with the Lorentz curvature
$R^{L}=\dr\go^L+\go^L\wedge\go^L$. Then  local Lorentz symmetry demands
that the Lorentz connection
$\go^L$ does not enter the r.h.s. of (\ref{lorunf}). Practically,
 it is more convenient to work with
\eqref{unf}  than with \eqref{lorunf} since $(D^{L})^2\neq
0$ while $\dr^2=0$. It is therefore necessary to
check if the unfolded equations \eqref{unf} admit a field
redefinition to \eqref{lorunf} to respect the local Lorentz covariance.
That  such a field redefinition is indeed available for HS systems
in all orders in interactions has been shown in \cite{properties, Vasiliev:Rev}
(see also \cite{SS:an}).

Recently,  an extension of $4d$ HS equations
 that contains HS invariant functionals has been
proposed in \cite{Vasiliev:1504}. One of such functionals conjectured
 to give an on-shell action is anticipated to play a crucial
role for the purposes of HS $AdS/CFT$ industry (see
\cite{KP}-\cite{GKT} for the incomplete list of references). In
view of absence of conventional fully nonlinear HS action
principle (see however \cite{BS}) the possibility of calculating
the on-shell action paves a way to testing HS $AdS/CFT$ proposal
explicitly at least at tree level. (A further extension accounting
for quantum corrections was also discussed in
\cite{Vasiliev:1504}.) The extension of \cite{Vasiliev:1504} is
designed to include higher-degree differential forms in addition
to the  zero- and one-forms of the original  HS theory. These
higher forms carry no local physical degrees of freedom being
expressed in terms of physical fields on mass shell. The
space-time four-form valued in the center of the HS algebra
corresponds to the density of the aforementioned on-shell action.

To calculate the HS invariant density  and compare it with the correlation
functions of the dual theory in the boundary limit is a
challenging technical problem even to the lowest orders. The main
complication is due to the involved nature of HS perturbation theory
despite considerable progress made in \cite{DMV}.
A related difficulty is the fact that
the extended HS equations have been written down in form \eqref{unf}
rather than in \eqref{lorunf} leading to non-covariant and highly
involved expressions for the invariant functional requiring
further field redefinition for the covariantization.

The existence of Lorentz covariant frame for the extended HS
equations was neither presented nor proven in the literature. In
this paper we fill in this gap providing the explicit Lorentz
covariant extended HS equations in four dimensions. One of the
main difficulties in obtaining these equations is the fact that
the Lorentz covariant differential $D^L$ does not admit
star-product realization in its Lorentz connection part. This
makes it hard finding field redefinition that lead to the
covariant form of equations. Instead of looking for such a field
redefinition, our strategy is to start from the part of the
equations for space-time zero-forms which can be straightforwardly
covariantized. Consistency of these equations imposes certain
restrictions on the form of the rest of the equations still not
constraining them entirely. At this stage the analysis requires a
bit of a guess work. An important  observation is that the very
requirement of local Lorentz symmetry restricts the form of
field-independent central terms in the equations that determine
the structure of interactions, which otherwise are fixed by rather
indirect functional class arguments and not by mere consistency of
HS equations. The availability of a Lorentz frame largely relies
on the existence of field-dependent Lorentz generators constructed
out of purely twistor sector of the extended HS equations thanks
to their specific form that admits interpretation in terms of a
deformed oscillator algebra generalizing the deformed oscillator
realization of the Lorentz algebra underlying usual HS equations
of \cite{more} (see also \cite{Vasiliev:Rev} and references
therein).

Lorentz covariant equations proposed in this paper contain an additional Lorentz connection on the
top of the tower of HS gauge fields. Their
integrability results in an overdetermined set of constraints
which are however fulfilled as a consequence of
spinor Fierz (Schouten) identities. To see that the proposed equations are
dynamically equivalent to those of
\cite{Vasiliev:1504} we observe additional Stueckelberg symmetry
generalizing that   found in \cite{IS} for the usual bosonic HS equations,
which allows one to set an additional Lorentz connection to zero thus reducing
equations to their ordinary unfolded frame yet providing explicit
field redefinition relating the two frames. Naively, Stueckelberg
nature of the spin-two Lorentz connection in HS master field implies that it is
gauge equivalent to zero. Once however we insist on the absence of the Lorentz
connection type contribution to the spin two sector within HS
master field, then the additional Lorentz connection gets perturbatively
determined up to true HS gauge symmetry transformations.

From technical standpoint the advantage of the proposed
covariant equations   as compared to those in
\cite{Vasiliev:1504} is due to a remarkable simplification of
perturbative series operators they deliver along the lines of
\cite{DMV}. To show this we reconsider perturbative expansion
about proper HS vacuum solution within the Lorentz covariant
approach and rederive operators which determine perturbative analysis of
the equations of motion. Our result reproduces formulas of
\cite{DMV} with the difference that they no longer contain Lorentz
connection, leading to dramatic simplification especially
for the twisted-adjoint case.

The paper is organized as follows. In section \ref{HScov} we
recall HS equations in four dimensions and explain a way of making
them explicitly Lorentz covariant suitable for the HS system
containing the  invariant functional. Then, in section \ref{excovHS} we
address a problem of covariantization of the extended  HS
equations. To do that we discuss a generalization of the deformed
oscillator algebra in section \ref{Genosc}. A manifestly covariant form
of HS equations is given in section \ref{Loreq}. Finally, in
section \ref{covth} we elaborate a covariant perturbation theory
for the obtained equations. Brief conclusions are in section \ref{con}.

\section{Lorentz covariant HS equations}\label{HScov}
Let us start by reviewing the HS equations in four
dimensions of \cite{more} (see also \cite{Vasiliev:Rev}) and their Lorentz
covariantization. The equations have the following standard form
\begin{align}
&\dr\W+\W*\W=i\theta^{A}\theta_{A}+i\eta B*\gga+i\bar{\eta}
B*\bar{\gga}
\,,\label{hs1}\\
&\dr B+[\W, B]_*=0\,,\label{hs2}
\end{align}
where $\W(Z,Y; K |x)$ is the 1-form in the double graded space
spanned by anticommuting $\dr x^{\underline{m}}$ and auxiliary
$\theta^A$ differentials. $\W$ contains HS potentials with all
their descendants in space-time subsector $W$ and the
compensator-like field $S$ in the $\theta$-subsector,
\begin{align}
&\W(Z,Y; K |x)=W_{\underline{m}}(Z,Y; K |x)\dr
x^{\underline{m}}+S_{A}(Z,Y; K |x)\theta^A\,,\\
&\{\dr x^{\underline{m}}, \dr x^{\underline{n}}\}=\{\dr
x^{\underline{m}}, \theta^A\}=\{\theta^A, \theta^B\}=0\,.
\end{align}
Space-time indices $\underline{m}, \underline{n}$ as well as
Majorana spinor fiber indices $A, B$ range four values. $B(Z,Y; K
|x)$ is a  0-form master field, containing lower spin
matter fields and HS curvatures.

Apart from $x$-dependence, $\W$ and $B$ depend on a number of
generating variables. Commuting twistor-like variables
$Y_{A}=(y_{\al}, \bar{y}_{\dal})$ and $Z_{A}=(z_{\al}
,\bar{z}_{\dal})$ are designed to pack up HS fields and auxiliary
fields, where spinor indices $\al, \gb,...$ range two values. The
associative star-product operation acts on functions
$f(Z,Y)$
\be\label{star}
(f*g)(Z,Y)=\ff{1}{(2\pi)^4}\int dU dV f(Z+U, Y+U)g(Z-V,
Y+V)e^{iU_A V^A}\,,
\ee
where $U_AV^A:=U_A V_B \gep^{AB}$ with some $sp(4)$-invariant symplectic form
$\gep_{AB}=-\gep_{BA}$. Indices are raised and lowered with the
aid of $\gep_{AB}$ as follows, $X^{A}=\gep^{AB}X_{B}$ and
$X_{A}=\gep_{BA}X^{B}$. The star product can be seen to induce the
following commutation relations
\be
[Y_A ,Y_B]_*=-[Z_{A}, Z_{B}]_*=2i\gep_{AB}\,,\qquad [Y_A,
Z_B]_*=0\,.
\ee
It admits the inner Klein operators
\be
\gk=e^{iz_{\al}y^{\al}}\,,\qquad \bar\gk=e^{i\bar z_{\dal}\bar
y^{\dal}}\,.
\ee
Their characteristic properties are
\be\label{Klpr}
\{\gk, y_{\al}\}_*=\{\gk, z_{\al}\}_*=0\,,\qquad \gk*\gk=1\,,
\ee
analogously in the antiholomorphic sector for $\bar{\gk}$.

To distinguish between the adjoint representation for HS
potentials and the twisted-adjoint for HS curvatures as well as to
have a room for topological degrees of freedom in HS system one
introduces Clifford variables $K=(k, \bar k)$ which are called the
outer Klein operators. $k$ anticommutes with the holomorphic
variables $y_{\al}, z_{\al}, \theta_{\al}$ and commutes with every
antiholomorhic one. Note that being similar to star-product
realized $\gk$, $k$ nevertheless  admits no
algebraic realization due to its anticommutativity with
$\theta^{\al}$. Analogously, $\bar k$ anticommutes with
antiholomorhic variables and commutes with all of the rest. In
addition,
\be
k^2=\bar k^2=1\,,\qquad [k, \bar k]=0.
\ee
The latter relation makes the dependence on Klein operators at
most bilinear
\be
\W=\sum_{i,j=0,1}\W_{i,j}k^i\bar k^{j}\,,\qquad
B=\sum_{i,j=0,1}B_{i,j}k^i\bar k^{j}\,.
\ee
Physical fields are encoded in $\W(-k,-\bar k)=\W(k, \bar k)$ and
$B(-k,-\bar k)=-B(k, \bar k)$. The rest decompose into infinite tower
of fields, each carrying at most a finite number of
dynamical degrees of freedom, being therefore topological.

There are two central elements that enter the r.h.s. of
\eqref{hs1}. One is $\theta^ A\theta_A$ which obviously
commutes with any variable. Another one is
\be
\gga=2k\gk\gd^{2}(\theta)\,,\qquad
\gd^2(\theta):=\ff12\theta^{\al}\theta_{\al}\,,
\ee
which commutes with everything including $\theta^{\al}$ thanks to
the Grassmann $\delta$-function, $\theta^{\al}\gd^{2}(\theta)=0$.
Analogously defined is $\bar\gga$. Finally, complex phase
$\eta=e^{i\phi}$ is the only free parameter of HS theory in four
dimensions with the r.h.s. of \eqref{hs1} linear in $B$.
For parity preserving cases when $\phi=0$ the theory is known as
the A--model and $\phi=\pi/2$ corresponds to the B--model
\cite{SS}.

In terms of $W$ and $S$ components of $\W$ \eqref{hs1}, \eqref{hs2}
can be rewritten as follows
\begin{align}
&\dr W+W*W=0\,,\label{HS1}\\
&\dr S+[W,S]_*=0\,,\label{HS2}\\
&\dr B+[W,B]_*=0\,,\label{HS3}\\
&S*S=-i\theta_{\al}\wedge \theta^{\al}(1+\eta B*k\gk)-
i\bar\theta_{\dal}\wedge \bar\theta^{\dal}(1+\bar\eta B*\bar k\bar \gk)\,,\label{HS4}\\
&[S,B]_*=0\,.\label{HS5}
\end{align}

To proceed to Lorentz covariance let us first recall that Lorentz covariant
derivative of a  Lorentz multispinor
$\phi_{\al_{1}\dots\al_{n}, \dal_{1}\dots\dal_{m}}$ is
\be\label{DL}
D^{L}\phi_{\al(n), \dal(m)}=\dr \phi_{\al(n),
\dal(m)}-n\go_{\al}{}^{\gb}\phi_{\gb\al(n-1),
\dal(m)}-m\bar{\go}_{\dal}{}^{\dgb}\phi_{\al(n), \dgb\dal(m-1)}\,,
\ee
where $\go_{\al\gb}$ and $\bar{\go}_{\dal\dgb}$ are holomorphic
and antiholomorphic parts of the Lorentz connection
$\go^{AB}=(\go_{\al\gb}, \bar{\go}_{\dal\dgb})$ which
generates Lorentz curvature 2-form via
\begin{align}
&(D^{L})^2\phi_{\al(n),
\dal(m)}=-nR_{\al}{}^{\gb}\phi_{\gb\al(n-1),
\dal(m)}-m\bar{R}_{\dal}{}^{\dgb}\phi_{\al(n), \dgb\dal(m-1)}\,,\label{D1}\\
&R_{\al\gb}=\dr\go_{\al\gb}-\go_{\al}{}^{\gga}\wedge\go_{\gga\gb}\,,\qquad
\bar
R_{\dal\dgb}=\dr\bar{\go}_{\dal\dgb}-\bar{\go}_{\dal}{}^{\dgga}\wedge\bar{\go}_{\dgga\dgb}\,,\\
&D^L R_{\al\gb}=D^L\bar{R}_{\dal\dgb}=0\,.
\end{align}
In HS equations \eqref{HS1}-\eqref{HS5} all spin-tensors are
packed into generating functions $W$, $S$ and $B$ with the aid of
the variables $Y^A$, $Z^A$ and also $\theta^{A}$ in case of $S$--field.
 In these terms, \eqref{DL}  can be rewritten   as
\be\label{DLop}
D^{L}f(Z,Y;\theta)=\left(\dr+\go^{AB}\left(Z_{A}\ff{\p}{\p
Z^B}+Y_{A}\ff{\p}{\p
Y^B}+\theta_{A}\ff{\p}{\p\theta^B}\right)\right)f(Z,Y;\theta)\,.
\ee
Note that the differential with respect to $Y$ and $Z$
operators in \eqref{DLop} can be realized in terms of star product
\begin{align}
&\go^{AB}\left(Z_{A}\ff{\p}{\p Z^B}+Y_{A}\ff{\p}{\p
Y^B}\right)=\go^{AB}[L_{AB}, f]_{*}\,,\\
&L_{AB}=-\ff i4\left(Y_AY_B-Z_AZ_B\right)\,.\label{L0}
\end{align}
Clearly, $L_{\al\gb}$ and $\bar{L}_{\dal\dgb}$ form $sl_2(\mathbb{C})$ Lorentz subalgebra. It follows then that if it were not for
$\theta$-dependence in $S$ field the local Lorentz symmetry on HS
equations could be restored by a simple field redefinition $W\to
W+\go^{AB}L_{AB}$. Indeed, in this case the de-Rham differential
$\dr$ in \eqref{HS2} and \eqref{HS3} turns into the Lorentz
differential $D^{L}$ by the $\go^{AB}L_{AB}$ shift. However
this shift does not act properly on $S$ field in \eqref{HS2} which
being $\theta$--dependent carries extra Lorentz indices
$S=S_{\al}\theta^{\al}+\bar{S}_{\dal}\bar\theta^{\dal}$
not rotated by $L_{AB}$. This is where the last term in
\eqref{DLop} becomes important. Since
$\theta_A\ff{\p}{\p\theta^B}$ cannot be realized via star-product
there must be a special reason for the local Lorentz
symmetry to take place\footnote{ A naive way out is to extend star
product spanned by bosonic $Y$ and $Z$ variables by adding
Clifford elements $\theta$. Such an extension while making scheme
explicitly Lorentz covariant appears to change the content of the
equations.}. As shown in \cite{properties, Vasiliev:Rev}
Lorentz covariance follows from the
property of the deformed oscillator algebra as we now recall.

For the sake of presentation simplicity, consider for a moment bosonic
truncation with no topological degrees of freedom in
\eqref{HS1}-\eqref{HS5}. General case is fully captured by this
example. We are interested in the $\theta$-sector, i.e. the sector of
space-time 0-forms to be referred to as the twistor sector.
Corresponding equations are
\begin{align}
&[S_{\al}, S_{\gb}]_*=-2i\gep_{\al\gb}(1+\eta B*\gk)\,,\label{deform1}\\
&\{S_{\al}, B*\gk\}_*=0\,,\label{deform2}
\end{align}
(similarly in the dotted sector) where fields $S$ and $B$ are now
$k$ and $\bar k$ independent. Equations
\eqref{deform1}-\eqref{deform2} have the form of deformed
oscillator algebra \cite{Vasiliev:1989qh,Vasiliev:1989re} where it
was observed that it  respects $sp(2)$ symmetry. Namely, the
generators
\be\label{Lordef}
M_{\al\gb}=\ff {i}{8}\{S_{\al}, S_{\gb}\}_*
\ee
form $sp(2)$  for any $B$,
\be
[M_{\al\al}, M_{\gb\gb}]_*=2\gep_{\al\gb}M_{\al\gb}\,.
\ee
Together with the generators in the antiholomorphic sector it
restores the full Lorentz algebra $sl_2(\mathbb{C})$.
Particularly, by \eqref{deform1}, \eqref{Lordef} these generators
Lorentz rotate the field $S$ itself
\be\label{Srot}
[M_{\al\al}, S_{\gb}]_*=\gep_{\al\gb}S_{\al}\,.
\ee
Since \eqref{Srot} mimics the action of the $\theta$--term in
\eqref{DLop} this suggests that the field-dependent
generators $M_{\al\gb}$ and $\bar{M}_{\dal\dgb}$ can underly
 the field redefinition that restores local Lorentz symmetry.
Indeed, at the nonlinear level the Lorentz
algebra action consists of two parts \eqref{L0} and \eqref{Srot}
suggesting the following field redefinition $W\to
W+\go^{AB}(L_{AB}+M_{AB})$. That this is indeed the case can be
checked directly (see \cite{Vasiliev:Rev,SS:an}). We however take
somewhat different way
which is most useful in the more complicated
extended HS system.

Let us assume that there exists a field redefinition that makes
local Lorentz symmetry manifest. Since such redefinition contains
Lorentz connection it cannot affect space-time 0-forms $S$ and
$B$. This implies that our bosonic truncated equations
\eqref{deform1} and \eqref{deform2} remain undeformed. The only
field that should be redefined is the space-time 1-form $W$. The
effect of the field redefinition $W\to W'=W+\go^{AB}(\dots)$ on
the zero-curvature conditions
\begin{align}
&\dr S_{\al}+[W,S_{\al}]_*=0\,,\\
&\dr B+W*B-B*\pi(W)=0\,,
\end{align}
where $\pi(y,\bar y, z, \bar z)=(-y,\bar y, -z, \bar z)$ will be
their covariantization of the form
\begin{align}
&D^L S_{\al}+[W',S_{\al}]_*=0\,,\label{Scov}\\
&D^{L} B+W'*B-B*\pi(W')=0\,\label{Bcov}
\end{align}
with $W'$ being free from the Lorentz connection.
Consistency \eqref{D1}
\be
(D^L)^2S_{\al}=R^{\gb\gga}[L_{\gb\gga},S_{\al}]_*-R_{\al}{}^{\gb}S_{\gb}
\ee
applied to \eqref{Scov} implies using \eqref{Lordef} that
\be [D^L
W'+W'*W'+R^{\gb\gga}(L_{\gb\gga}-\ff i4 S_{\gb}*S_{\gga}),
S_{\al}]_*=0
\ee
suggesting eventually the following Lorentz covariant system
\begin{align}
&D^L W'+W'*W'+R^{AB}\left(L_{AB}-\ff i4
S_{A}*S_{B}\right)=0\,,\label{blor1}\\
&R^{AB}:=\dr \go ^{AB}-\go ^{AC}\go_{C}{}^{B}\,,\\
&D^L S_{A}+[W',S_{A}]_*=0\,,\\
&D^{L} B+W'*B-B*\pi(W')=0\,,\\
&[S_{\al}, S_{\gb}]_*=-2i\gep_{\al\gb}(1+\eta B*\gk)\,,\qquad
[\bar S_{\dal}, \bar S_{\dgb}]_*=-2i\gep_{\dal\dgb}(1+\bar \eta B*\bar \gk)\,,\\
&\{S_{\al}, B*\gk\}_*=0\,,\qquad \{\bar S_{\dal},
B*\bar\gk\}_*=0\,.\label{blor5}
\end{align}
The obtained equations are consistent and exhibit manifest local
Lorentz symmetry. In this form they were written down in
\cite{SS:an}. By the field redefinition
\be
W'=W-\go^{AB}\left(L_{AB}-\ff i4 S_A*S_B\right)
\ee
 \eqref{blor1}-\eqref{blor5} is reduced down to the bosonic version of
\eqref{HS1}-\eqref{HS5}.

Before we pursue covariantization of general case
\eqref{hs1}, \eqref{hs2} let us make few comments. First, we see
that twistor sector of HS equations plays crucial role for local
Lorentz symmetry. Namely, the possibility to write down covariant
equations through a field redefinition relies on the existence of
field dependent Lorentz generators built from the on-shell twistor
fields (see \eqref{Srot}). For example, while one can relax
\eqref{hs1} by dropping off $\theta^A\theta_A$ term still keeping
equations consistent\footnote{This example is physics wise
unappropriate due to its failure of reproducing free Fronsdal
equations.} there would be no local Lorentz symmetry as we will
see. We will keep this strategy of examining twistor sector in our
analysis of  more complicated HS extended equations.

Second, the obtained equations \eqref{blor1}-\eqref{blor5} do not
determine Lorentz connection $\go_{AB}$ by itself. As shown in
\cite{IS} where the Lorentz covariant equations were used,
this system possesses St\"{u}ckelberg symmetry that allows one to
gauge away the connection. The connection nonetheless can be
perturbatively fixed up to HS gauge transformations once the field
$W'$ is demanded to be free of
 connection-type components. This implies that the physical
 Lorentz connection given by $\go_{AB}$ is unique. To be more
specific, normally one associates physical fields with
cohomologies represented by $Z$-independent functions (see
e.g., \cite{Vasiliev:Rev}, \cite{SS:an}). Therefore we need no
(anti)holomorphic bilinears to reside in $W'$ which if present
would correspond to yet another connection-type fields. Hence, we
set
\be
\ff{\p^2}{\p y^{\al}\p y^{\gb}}W'\Big|_{Z,Y=0}= \ff{\p^2}{\p \bar
y^{\dal}\p \bar y^{\dgb}}W'\Big|_{Z,Y=0}=0\,.
\ee

Now let us consider generic case of HS equations that contain
double set of physical fields along with the topological ones in
accordance with \eqref{hs1}-\eqref{hs2}. To make them Lorentz
covariant we use the following trick. We note that the $S$-term in
parentheses of \eqref{blor1} can be formally represented as
$\ff{\p}{\p\theta^A}\W$. This operator however does not respect
the chain rule  on a product of fields due to their dependence on
outer Klein operators $k$ and $\bar k$. To fix that we introduce
an additional operator $\rho$  that anticommutes with $k$ and
commutes with the rest (similar operator appears in $3d$ model of
\cite{Prokushkin:1998bq}) and analogously we introduce $\bar\rho$
that anticommutes with $\bar k$
\begin{align}
\rho k+k\rho=0\,,\qquad \rho\theta-\theta\rho=0\,,\qquad
\rho^2=1\,.
\end{align}
These properly account for sign flip of the outer Klein operators
$k$ and $\bar k$ in their master field dependence allowing to
define derivations
\be
\p_{\al}:=\rho\ff{\p}{\p\theta^{\al}}\,,\qquad
\p_{\dal}:=\bar\rho\ff{\p}{\p\bar\theta^{\dal}}
\ee
respecting the chain rule
\begin{align}
\p_{A}(F^{(p)}*G^{(q)})=\p_{A}F^{(p)}*G^{(q)}+(-)^p F^{(p)}*\p_{A}
G^{(q)}\,\label{chain}
\end{align}
and allowing one to extend the bosonic case to the following
general Lorentz covariant HS equations
\begin{align}
&D^{L}\W+\W*\W+R^{AB}\left(L_{AB}-\ff{i}{4}\p_A\W*\p_B\W\right)=i\theta^{A}\theta_{A}+i\eta
B*\gga+i\bar{\eta} B*\bar{\gga}\,,\label{cov_hs1}\\
&D^{L}B+[\W, B]=0\,.\label{cov_hs2}
\end{align}
Note that the dependence on $\rho$ and $\bar\rho$ drops out from
\eqref{cov_hs1} upon generating certain sign factors. The consistency of
\eqref{cov_hs1}, \eqref{cov_hs2} is easy to check using
\begin{align}
&(D^L)^2 f(Z,Y;
\theta)=R^{AB}[L_{AB},f]+R^{AB}\theta_{A}\ff{\p}{\p\theta^B}
f\,,\label{D2}\\
&D^{L}(R^{AB}L_{AB})=0\,.
\end{align}
For $\go_{AB}=0$ one gets back to \eqref{hs1}, \eqref{hs2}. For
$\go_{AB}\neq 0$ following \cite{IS} we observe an extra
St\"{u}ckelberg symmetry
\begin{align}
&\gd_{\xi}\go_{AB}=\xi_{AB}\,,\qquad \gd_{\xi}B=0\,,\\
&\gd_{\xi}\W=-\xi^{AB}\left(L_{AB}-\ff
i4\p_{A}\W*\p_{B}\W\right)\,.
\end{align}

It is interesting to note, that by simple rescaling of
$\theta$-variables along with master fields one can modify the
first term on the r.h.s. of \eqref{hs1} to be
$i\nu\theta^A\theta_A$, where $\nu$ is  an arbitrary real
number. This rescaling correspondingly modifies covariant equation
\eqref{cov_hs1}
\be
D^{L}\W+\W*\W+R^{AB}\left(L_{AB}-\ff{i}{4\nu}\p_A\W*\p_B\W\right)=i\nu\theta^{A}\theta_{A}+i\eta
B*\gga+i\bar{\eta} B*\bar{\gga}\,,
\ee
which makes the limit $\nu\to 0$ meaningless  unlike the same limit for
the noncovariant system. Therefore, the very requirement of Lorentz
symmetry rules out some formally consistent systems.
 This particular limit can be ruled out by comparing
the free equations it provides to the proper free HS ones though.
In other words \eqref{cov_hs1} written down in the
non-covariant frame $R^{AB}=0$ does not reproduce proper free HS
equations upon linearization about $AdS$ background for $\nu=0$.

\section{Covariantization of extended higher-spin system}\label{excovHS}
\subsection{Lorentz covariance in the twistor sector}
An extended version of HS equations in
four dimensions was proposed in \cite{Vasiliev:1504}. Being physically
equivalent to the original HS equations it contains higher
differential forms which are expressed on-shell via dynamical HS
fields. This extension is aimed at generating a gauge-invariant
functional represented  as a space-time integral over a
four-form and is organized as follows.

The one-form $\W$ in  $(\dr x\oplus\theta)$-differentials is
extended to three-forms $\dr x^{n}\wedge\theta^{3-n}\,,n=0,\dots,
3$. Analogously, the zero-form $B$ \eqref{hs2} gets enhanced to
all two-forms $\dr x^{n}\wedge\theta^{2-n}\,,n=0,\dots, 2$. This
way one introduces odd forms $\W$ and even forms $\B$. In
different context a similar extension was also given in \cite{BS}.
To let higher forms be nontrivially expressed in terms of
dynamical HS fields of the original system, equations \eqref{hs1},
\eqref{hs2} have to be properly modified since otherwise the
higher forms would receive no non-zero source. The nontrivial HS
dynamics in the original system rests on the sources bilinear in
$\theta$ on the r.h.s. of \eqref{hs1} for the twistor field $S$.
Now one has to introduce a source for a new twistor field resided
in $\W$ which is a $\theta^3$-form. Compatibility of the modified
equations demands any such source be central. Among few of such
elements $g\gga*\bar\gga$, where $g$ is a constant,
 was argued in \cite{Vasiliev:1504} to be the proper one. Eventually,
the proposed equations are
\begin{align}
&\dr\W+\W*\W=i\theta^{A}\theta_{A}+i\eta \B*\gga+i\bar{\eta}
\B*\bar{\gga}+ig\gga*\bar{\gga}+\L
\,,\label{ehs1}\\
&\dr \B+[\W, \B]_*=0\,.\label{ehs2}
\end{align}
Here $\L$ is $Y$, $Z$, $k$, $\bar k$ and $\theta$--independent
pure space-time four-form to be associated with the invariant
functional density\footnote{Equivalently, one could introduce no
$\L$ on the r.h.s. of \eqref{ehs1} instead associating the
four-form $(\W*\W)|_{Z,Y,\theta=0}$ with the invariant functional
density.}. Equation \eqref{ehs1} may be deformed to contain
$F_*(\B)$ on the r.h.s. instead of  $\B$ and $G(\B)$ instead of
$g$. We took $F=\eta\B$, $G(\B)=g$ for simplicity. The system is
invariant under the following gauge transformation
\begin{align}
&\gd\W=\dr\gep+[\W,\gep]_*+i\eta\xi*\gga+i\bar\eta\xi*\bar\gga\,,\label{g1}\\
&\gd\B=\dr\xi+\{\W,\xi\}_*+[\B,\gep]_*\,.\label{g2}
\end{align}

Our goal is to prove that \eqref{ehs1}, \eqref{ehs2} respects
Lorentz symmetry, rewriting it in a
manifestly Lorentz covariant form. A major asset on this way is
the existence of Lorentz generators in the twistor sector of
\eqref{ehs1}, \eqref{ehs2} as we now show.

Let us introduce components for the twistor fields as follows
\begin{align}
&\W\Big|_{\dr
x=0}=s_{A}\theta^A+2t_{\al}\theta^{\al}\gd^{2}(\bar\theta)+2\bar
t_{\dal}\bar\theta^{\dal}\gd^{2}(\theta)\,,\\
&\B\Big|_{\dr x=0}=B+2b\gd^2(\theta)+2\bar
b\gd^2(\bar\theta)+b_{\al\dal}\theta^{\al}\bar{\theta}^{\dal}\,.
\end{align}
Assuming for brevity the bosonic truncation with $k\bar k=1$ from
 the $\theta$--sector of
\eqref{ehs1}, \eqref{ehs2} one extracts the following algebraic
constraints
\begin{align}
&[s_{\al}, s_{\gb}]_*=-2i\gep_{\al\gb}(1+\eta B*\gk)\,,\qquad [s_{\al}, \bar{s}_{\dal}]_*=0\,,\label{tw1}\\
&[t_{\al}, s^{\al}]_*+[\bar{t}_{\dal},
\bar{s}^{\dal}]_*=-2i(\eta\bar{b}*\gk+\bar{\eta}b*\bar\gk+g\gk*\bar\gk)\,,\label{tw2}\\
&s_{\al}*B+B*\pi(s_{\al})=0\,,\label{tw3}\\
&s_{\al}*\bar b-\bar
b*\bar{\pi}(s_{\al})+t_{\al}*B+B*\pi(t_{\al})=0\,,\label{tw4}\\
&\ff12\left(s_{\al}*b^{\al}{}_{\dal}+b^{\al}{}_{\dal}*\pi(s_{\al})\right)+\bar{t}_{\dal}*B-B*\pi(\bar{t}_{\dal})=0\,.\label{tw5}
\end{align}
An important observation is that equations \eqref{tw1}-\eqref{tw5}
admit Lorentz symmetry which can be realized as gauge
transformation \eqref{g1}. This can be seen most easily for
$\B=0$. Indeed from the gauge transformations
\begin{align}
&\gd_{\Lambda}s_{\al}=[s_{\al}, \gep]_*\,,\\
&\gd_{\Lambda} t_{\al}=[t_{\al}, \gep]_*+[\phi,
s_{\al}]+[\psi_{\al}{}^{\dgb}, \bar s_{\dgb}]\,,\label{tr2}\\
&\gd_{\Lambda}\bar t_{\dal}=[\bar t_{\dal},
\gep]_*+[\bar\phi, \bar
s_{\dal}]-[\bar\psi^{\gb}{}_{\dal}, s_{\gb}]\,,\label{tr3}
\end{align}
where $\gep$, $\phi$ and $\psi_{\al\dal}$ are some gauge
parameters, it follows that
$\gd_{\Lambda}s_{\al}=\Lambda_{\al}{}^{\gb}s_{\gb}$ and
$\gd_{\Lambda}t_{\al}=\Lambda_{\al}{}^{\gb}t_{\gb}$ provided that
\be\label{defLor}
\gep=-\ff{i}{4}\Lambda^{\al\gb}s_{\al}*s_{\gb}\,,\qquad
\phi=\ff{i}{4}\Lambda^{\al\gb}\{s_{\al}, t_{\gb}\}_*\,,\qquad
\psi_{\al\dal}=-\ff{i}{4}\Lambda_{\al}{}^{\gb}\{s_{\gb}, \bar
t_{\dal}\}_*\,.
\ee
This fact suggests that the whole system \eqref{ehs1},
\eqref{ehs2} should be  Lorentz covariant  upon an appropriate
field redefinition.

\subsection{Generalized deformed oscillator algebra}\label{Genosc}
Analogously to the original construction \eqref{deform1},
\eqref{deform2} that underlies nonlinear HS equations, Eqs.
\eqref{tw1}-\eqref{tw5} give rise to a generalization of the
deformed oscillator algebra.

Standard oscillator (Weyl) algebra is the enveloping algebra of the relations
\be\label{Weyl}
[y_{\al}, y_{\gb}]=-2i\gep_{\al\gb}\,.
\ee
It can be  extended by adding an outer element $K$
anticommuting with $y_{\al}$ which is called Klein operator
\be\label{K}
\{y_{\al}, K\}=0\,,\qquad KK=1.
\ee
With this new element there exists a one-parametric deformation of
the algebra
\be\label{def}
[y_{\al}, y_{\gb}]=-2i\gep_{\al\gb}(1+\nu K)\,,
\ee
where $\nu$ is an arbitrary constant (central element). Jacobi
relations
\be
[[y_{\al}, y_{\gb}], y_{\gga}]+cycle=0
\ee
automatically hold for \eqref{def} provided that indices $\al, \gb,
\gga$ range two values. As was stressed already in \eqref{Lordef}
the remarkable property of \eqref{def} is that for any $\nu$ the
bilinears of $y$ generate $sp(2)$
\be
\label{M}
M_{\al\gb}=\{y_{\al}, y_{\gb}\}\,,\qquad [M_{\al\al},
M_{\gb\gb}]\sim\gep_{\al\gb}M_{\al\gb}\,.
\ee
Adding another copy of deformed oscillator $\bar y_{\dal}$ in the
dotted sector  one extends this $sp(2)$ to the four dimensional
Lorentz algebra $sp(2|\mathbb{C})$. Eqs.~\eqref{tw1}-\eqref{tw5}
provide a generalization of \eqref{def} that still respects the
$sp(2|\mathbb{C})$ symmetry. The simplest such an extension arises
if one sets $B=const$ and $b=0$ in which case the defining
relations become
\begin{align}
&[s_{\al}, s_{\gb}]=-2i\gep_{\al\gb}(1+\nu K)\,,\qquad &&[\bar
s_{\dal}, \bar s_{\dgb}]=-2i\gep_{\dal\dgb}(1+\nu\bar K)\,,\label{eq1}\\
&\{s_{\al}, K\}=0\,,\qquad &&\{\bar s_{\dal}, \bar K\}=0\,,\\
&\{t_{\al}, K\}=0\,,\qquad &&\{\bar t_{\dal}, \bar K\}=0\,,\\
&[s_{\al}, \bar s_{\dal}]=0\,,\qquad &&[t_{\al}, s^{\al}]+[\bar
t_{\dal}, \bar s^{\dal}]=2i g K\bar K\,.\label{eq4}
\end{align}

Since the part of the $sp(2)\oplus sp(2)$ Lorentz transformations in
(\ref{tr2}) and (\ref{tr3}) associated with $\phi$, $\bar\phi$, $\psi$ and
$\bar\psi$ is generated by the gauge symmetries of the  higher differential form
fields, it can be argued that the Lorentz $sp(2)\oplus sp(2)$
symmetry is still generated by  $M_{\ga\gb}$ (\ref{M})
on the factor algebra of the universal enveloping algebra of
 \eqref{eq1}-\eqref{eq4} over the ideal generated by the higher-spin gauge
transformations.

\subsection{Lorentz covariant equations}\label{Loreq}
To find proper set of fields that brings \eqref{ehs1}, \eqref{ehs2}
into a manifestly Lorentz covariant form we again use the fact
that such a field redefinition containing Lorentz connection
$\go_{\al\gb}$, $\bar\go_{\dal\dgb}$ leaves twistor fields
(space-time 0-forms) unaffected. Therefore the covariantized
version for space-time evolution of these fields is reproduced by
replacing $\dr$ with $D^L$ in the original equations
\eqref{ehs1}, \eqref{ehs2}. These in turn impose further
restrictions on the other equations via integrability condition
\eqref{D2} still not fixing the latter entirely. Skipping technical details let us give the
final result for the manifestly Lorentz covariant HS equations
\begin{align}
&D^{L}\W+\W*\W+R^{AB}\left(L_{AB}-\ff{i}{4}\p_A\W*\p_B\W\right)=i\theta^{A}\theta_{A}+i\eta
\B*\gga+i\bar{\eta} \B*\bar{\gga}+ig\gga\bar{\gga}+\L-\notag\\
&-\ff{\eta}{4}R^{\al\gb}\p_{\al}\B*\p_{\gb}\gga-\ff{\bar{\eta}}{4}\bar{R}^{\dal\dgb}
\p_{\dal}\B*\p_{\dgb}\bar{\gga}+
\ff{i\eta}{32}R^{\al\al}R^{\gb\gb}\p_{\al}\p_{\gb}\B*\p_{\al}\p_{\gb}\gga
+
\ff{i\bar\eta}{32}\bar R^{\dal\dal}\bar R^{\dgb\dgb}\p_{\dal}\p_{\dgb}\B*\p_{\dal}\p_{\dgb}\bar\gga\,,\label{cov_ehs1}\\
&D^{L}\B+[\W, \B]-\ff{i}{4}R^{AB}\{\p_A\B,
\p_B\W\}_*=0\,.\label{cov_ehs2}
\end{align}
Equations \eqref{cov_ehs1}, \eqref{cov_ehs2} reproduce
\eqref{ehs1}, \eqref{ehs2} at $\go_{AB}=0$ in which case Lorentz
connection is contained in $\W$, being equivalent to the latter up
to a field redefinition at $\go_{AB}\neq 0$. This field
redefinition can be found in the form of Abelian infinitesimal
Stueckelberg transformation
\begin{align}
&\gd_{\xi}\go_{AB}=\xi_{AB}\,,\qquad \gd_{\xi}\B=\ff
i4\xi^{AB}\{\p_{A}\B, \p_{B}\W\}_{*}\,,\label{St1}\\
&\gd_{\xi}\W=-\xi^{\al\gb}(L_{\al\gb}-\ff
i4\p_{\al}\W*\p_{\gb}\W)-\ff{\eta}{4}\xi^{\al\gb}\p_{\al}\B*\p_{\gb}\gga
+\ff{i\eta}{32}(\xi^{\al\al}R^{\gb\gb}+\xi^{\gb\gb}R^{\al\al})\p_{\al}\p_{\gb}\B*\p_{\al}\p_{\gb}\gga+c.c.\label{St2}
\end{align}
To check consistency of \eqref{cov_ehs1}, \eqref{cov_ehs2} is a bit
tiresome and needs to use Fierz identities.

Let us briefly sketch the consistency check. Acting with $D^L$ on
eqs. \eqref{cov_ehs1}, \eqref{cov_ehs2} and using \eqref{D2} one
arrives at algebraic consequences graded by the degree of the
Lorentz curvature $R_{AB}$. One should bear in mind that among
different bilinears in fields entering these equations only those
actually contribute which do not exceed the differential form
degree of $D^L\W$ in \eqref{cov_ehs1} and of $D^L\B$ in
\eqref{cov_ehs2}. The rest should be discarded. Hence, one is left
with $O(R^0)$, $O(R)$ and $O(R\wedge R)$ consistency constraints
to be checked independently.

$O(R^0)$ constraints are automatically satisfied since they
correspond to the original system \eqref{ehs1}, \eqref{ehs2} which
is consistent. First nontrivial contribution coming from the
$O(R)$-terms gives
\be
\{\B*\p_{\al}\gga, \p_{\al}\W\}+[\p_{\al}\B*\p_{\al}\gga,
\W]-\p_{\al}[\W,\B]*\p_{\al}\gga=0\,,
\ee
 (similarly for the antiholomorphic part) which is equivalent to
\be\label{R1}
-\left(
\B*\p_{\al}\gga*\p_{\al}\W+\p_{\al}\B*\p_{\al}\gga*\W\right)=\left(
\p_{\al}\B*\W+\B*\p_{\al}\W\right)*\p_{\al}\gga\,.
\ee
Looking at \eqref{R1} we observe that the $\theta^2$-components of
$\W$ and $\B$ do not contribute because of $\theta^3\equiv 0$ and
symmetrization $\theta_{\al}\theta_{\al}\equiv 0$. That
$\theta_{\al}\theta_{\al}\equiv 0$ also implies that the terms
with both fields $\W$ and $\B$ being linear in $\theta$ do not
contribute. Therefore, only $\W$ linear in $\theta$ components in
the first term and $\theta$ -- independent in the second survive
on the l.h.s. of \eqref{R1}. Using
$\p_{\al}\gga*\W=\W(\theta,-\bar{\theta},
-dx)*\p_{\al}\gga=-\W*\p_{\al}\gga$ for $\theta$-independent
components and
$\p_{\al}\gga*\p_{\al}\W=-\p_{\al}\W(\theta,-\bar{\theta},
-dx)*\p_{\al}\gga=-\p_{\al}\W*\p_{\al}\gga$ for those linear in
$\theta$ we find that \eqref{R1} holds true.

For \eqref{cov_ehs2} the $O(R)$-consistency implies
\be
[\B,\p_{\al}\B*\p_{\al}\gga]=[\B*\p_{\al}\gga, \p_{\al}\B]\,,
\ee
which is also true by similar arguments.

Analyzing consistency of \eqref{cov_ehs1}, \eqref{cov_ehs2} up to
the $R^2$ level one has to remember that only differential forms
of degree $dx^4 \theta^A$ can show up in $(D^{L})^2\W+\dots$ and
$dx^4$ in $(D^{L})^2\B+\dots$. Indeed, the degree of $\W$ is
$(p,3-p)$, where $p=0,\dots, 3$ is attributed to $dx^p$.
Therefore, $\#(D^L)^2\W=(p+2, 3-p)$. Now, the top form $R\wedge R$
implies that $p=2$ and the only terms that should be analyzed
within consistency constraints are $dx^4 \theta^A$. The same
argument applies to $(D^L)^2\B+\dots$ equation. In practice this
has the result that only the twistor sector contribution to the
consistency relation has to be checked. This turns out to be
fulfilled for \eqref{cov_ehs1}, \eqref{cov_ehs2}. This analysis
implies in particular that had the system \eqref{cov_ehs1},
\eqref{cov_ehs2} contained higher differential forms it would not
be Lorentz covariant unless a proper deformation in the sector of
higher differential forms is taken into account.

Another interesting point is that as advertised the Lorentz
symmetry justifies the choice of the central element
$ig\gga*\bar\gga$ in \eqref{ehs1}. To see that let us denote it by
$c$.  Then the consistency of \eqref{cov_ehs1} implies in
particular
\be\label{cons}
R^{AB}\{\p_{A}c, \p_{B}\W\}_*=0\,,
\ee
which is true for $c=ig\gga*\bar\gga$ but {\it not} for other
$\theta^4$--central elements. For instance, for
$c=\theta^{\al}\theta_{\al}\bar\theta^{\dal}\bar\theta_{\dal}$,
\eqref{cons} is not true any longer. Therefore,
\eqref{cov_ehs1}, \eqref{cov_ehs2} are consistent only for the  specific
central element $g\gga*\bar{\gga}$  that governs the invariant
functional density $\L$.

In our analysis of the extended HS equations \eqref{ehs1},
\eqref{ehs2} we have chosen their interaction phase ambiguity to
be a field-independent complex number $\eta$. An extension of the
proposed Lorentz covariantization scheme to an arbitrary $F_*
(\B)$ is simple, resulting in minor modification of
\eqref{cov_ehs1}, \eqref{cov_ehs2}. Namely, to introduce
field-dependent phase in a Lorentz covariant way one simply
replaces $\eta\B$ with $F_* (\B)$, $\bar\eta \B$ with $\bar F_*
(\B)$ and $g$ with $G_* (\B)$ everywhere on the r.h.s. of
\eqref{cov_ehs1} while leaving \eqref{cov_ehs2} intact. Thus
modified equations turn out to remain consistent.

\section{Covariant perturbation theory}\label{covth}

Lorentz covariance of the proposed scheme is manifest provided  it is not explicitly broken
by the homotopy calculations in the perturbative analysis. This is true for
the vast class of Lorentz-covariant homotopies proposed recently in
\cite{Gelfond:2018vmi} where a subclass of homotopies leading to the local
vertices  was identified. That these homotopies properly reproduce the results
of \cite{Vasiliev:2016xui,Vasiliev:2017cae} is shown in \cite{DGKV}. In this paper
we  illustrate the usefulness of the proposed formulation
using the conventional homotopy allowing us to apply the approach
developed recently in \cite{DMV}.

Equations \eqref{ehs1}, \eqref{ehs2} naturally extend HS equations
\eqref{hs1}, \eqref{hs2} in a sense that all new higher-forms come
hand in hand with the structure of the original equations. A new
ingredient is the central element $ig\gga*\bar\gga$ giving rise to
the nontrivial invariant functional in perturbation theory. It
also delivers nontrivial vacuum value for $\W_0$ field which turns
out to be no longer polynomial in oscillators. However what really
complicates the perturbation theory is that the number of
elementary homotopy calculations within HS equations grows
combinatorially with the degree of differential forms and that the
naive result turns out  to be involved due to the terms containing
background Lorentz connection $\go_{AB}$.

As shown in \cite{DMV} one can dodge the first problem by finding
explicit expressions for operators originated from repeated
homotopy integration. The obtained formulas automatically account
combinatorial contributions of  terms coming from different
differential forms. As for the second problem, the reason why
noncovariant terms do show up is that the Lorentz covariance of
\eqref{ehs1}, \eqref{ehs2} is not manifest. These can be eliminated
using the field redefinition obtained in \eqref{St1}, \eqref{St2}.
However  a much simpler scheme developed in this section is to
reconsider the perturbation theory of \cite{DMV}
within the new Lorentz-covariant setup of \eqref{cov_ehs1}, \eqref{cov_ehs2}.

We start  with the proper vacuum solution to
\eqref{cov_ehs1}, \eqref{cov_ehs2} corresponding to
$AdS_4$ space-time,
\be\label{vac}
\W_{0}=\ff{i}{2}e^{\al\dal}y_{\al}\bar{y}_{\dal}+\theta^A
Z_A+\W_{0}^{3}\,,\qquad R^{\al\gb}_{0}=-e^{\al}{}_{\dgb}\wedge
e^{\gb\dgb}\,,\qquad \B_0=0\,,
\ee
where $e^{\al\dal}$ is the $AdS_4$ vierbein field
and $\W_0^3$ is the vacuum part of the 3-form which for now is
not important and will be specified later on. Using \eqref{vac} as a background in
perturbation theory we note that both \eqref{cov_ehs1} and
\eqref{cov_ehs2} lead to similar equations
\be\label{pert}
D^{L}f+[S_0, f]_*+[W_0, f]_*-\ff{i}{4}R^{AB}\{\p_{A}f,
\p_{B}S_0\}_*=J\,.
\ee
Here, $S_0=\theta^A Z_A$,
$W_0=\ff{i}{2}e^{\al\dal}y_{\al}\bar{y}_{\dal}$ and $f$ is either
$\W$ or $\B$, while  $J$  denotes the leftover terms appearing
in \eqref{cov_ehs1} or \eqref{cov_ehs2}. The dependence on extra
Klein operators in $f$ leads to different realization of the
commutator $[W_0, f]$ driving to the adjoint or twisted-adjoint
representations. It is important to note that all of the remaining
terms on the l.h.s. of \eqref{pert} act evenly as
differential operators in these two possible cases. Consider the
two cases in more detail.

\subsection{Adjoint case}

In this case $f$ is either independent of $k$ and $\bar k$ or
bilinear $f\to f k\bar k$. We separate the dependence on kleinians
so as to have $f$ Klein independent. One then has
\be\label{mainp}
\D f-2i\theta^A\ff{\p}{\p Z^A} f=J\,,
\ee
where
\begin{align}\label{D}
\D=&\dr+\go_{L}^{AB}\left( Y_{A}\ff{\p}{\p Y^B}+Z_{A}\ff{\p}{\p
Z^B}+\theta_{A}\ff{\p}{\p \theta^B}\right)-\ff{i}{2}R^{AB}\left(
Z_{A}\ff{\p}{\p\theta^B}+i\ff{\p}{\p Y^A}\ff{\p}{\p
\theta^B}\right)-\notag\\
&-e^{AB}\left( Y_{A}\ff{\p}{\p Y^B}-i\ff{\p}{\p Z^A}\ff{\p}{\p
Y^B}\right)\,.
\end{align}
$f$ can be found from \eqref{mainp} up to a purely gauge part as
\be\label{sol}
f=\Delta^{-1}J+g\,,\qquad \D g=\left(J-\D
f\right)\Big|_{Z=\theta=0}=:\mathcal{H}J\,,
\ee
where $g$ is $Z$-- and $\theta$--independent $\dr_Z$-cohomology part of
$\dr_Z:=\theta^A \frac{\p}{\p Z^A}$ (see
\cite{DMV} for more detail) and formal inversion of operator
$\Delta=\D-2i\dr_{Z}$ amounts to
\be\label{conhom}
\Delta^{-1}J=-\ff{1}{2i}\sum_{n=0}^{\infty}\left(
\ff{\dr^{*}_{Z}\D}{2i}\right)^{n}\dr^{*}_{Z}J\,,\qquad
\dr^{*}_{Z}J=Z^{A}\ff{\p}{\p\theta^A}\int_{0}^{1}\ff{dt}{t}J(tZ,
Y; t\theta)\,.
\ee
Since $(\dr^{*}_{Z})^2=0$, everything in $\D$ that commutes with
$\dr^{*}_{Z}$ brings no contribution to $\Delta^{-1}$. The last term
in the first line of \eqref{D} that contains curvature $R^{AB}$
while not commuting with $\dr^{*}_{Z}$ still adds no contribution to
$\Delta^{-1}$. This follows from $\dr^{*}_{Z}R^{AB}\left(
Z_{A}\ff{\p}{\p\theta^B}+i\ff{\p}{\p Y^A}\ff{\p}{\p
\theta^B}\right)\dr^{*}_{Z} J\equiv 0$. Hence, the only term that
matters in $\D$ is
\be
\D\sim ie^{AB}\ff{\p}{\p Z^A}\ff{\p}{\p Y^B}\,.
\ee
This way one finally gets
\begin{align}\label{ad}
\Delta^{-1}_{ad}J=&-\ff{1}{2i}Z^{A}\ff{\p}{\p\theta^A}\int_{0}^{1}\ff{dt}{t}
e^{\ff{1-t}{2t}e^{BC}\ff{\p^2}{\p Y^B\p\theta^C}}J(tZ, Y;
t\theta)=\notag\\
=&-\ff{1}{2i}Z^{A}\ff{\p}{\p\theta^A}\int_{0}^{1}\ff{dt}{t}
J\left(tZ,Y_A+\ff{1-t}{2t}e_{A}{}^{B}\ff{\p}{\p\theta^B};
t\theta\right)\,.
\end{align}
Substituting \eqref{ad} into the r.h.s. of the second
equation \eqref{sol} one finds the projection to cohomology
operator
\be\label{Had}
\mathcal{H}_{ad}J=J\left( 0,
Y_{A}+\ff{1}{2}e_{A}{}^{B}\ff{\p}{\p\theta^B};
\theta\right)\Big|_{\theta=0}\,.
\ee
Remarkably the $R$--curvature term does not enter perturbative
operators \eqref{ad}, \eqref{Had} at all as one reproduces
formulas from \cite{DMV} with the Lorentz connection set to zero.

A systematic way of obtaining explicit form of homotopy operators
which is most useful in twisted-adjoint case was proposed in
\cite{DMV} as we briefly recall. Consider a general equation of
the form
\be\label{gen_eq}
\Delta f:= \dr f + \D f=J\,,
\ee
where
\be
\Delta^2=0\,,\qquad \dr^2=0\,.
\ee
The consistency of \eqref{gen_eq} requires
\be\label{deltaJ_0}
\ \dr J + \D J=0\,.
\ee
Then, if a resolution of identity for $\dr$
\be\label{res_d}
\{\dr, \dr^{*}\}+h=Id\,
\ee
is known and
\be\label{d*D}
\{\dr^{*}, \D\}=0\,
\ee
holds true, a general solution to \eqref{gen_eq} can be written. To
this end one applies \eqref{res_d} to $J$ in \eqref{gen_eq} and,
making use of \eqref{deltaJ_0} and \eqref{d*D}, rewrites it as
\be\label{gen_eq2}
\ \dr(f-\dr^{*}J) + \D (f-\dr^{*}J) = hJ\,.
\ee
A general solution to \eqref{gen_eq2} is
\be\label{gen_sol}
\ f = \dr^{*}J+g+ \Delta \epsilon + \chi\,,
\ee
where an arbitrary function $\epsilon$ and general cohomology
$\chi\in H(\Delta)$ constitute a general solution to the homogeneous
equation, while $g$ solves
\be
\ h \D g = hJ\,.
\ee

This can be used to build a resolution of identity for the operator
from \eqref{mainp}. We have $\Delta_{ad} = \D_{ad} -2i \dr_{Z}$, where $\dr_{Z} =
\theta^A\ff{\p}{\p Z^A}$. For $\dr_{Z}$ one has
\begin{align}\label{d*}
\ &\dr^{*} f(Z;\theta) = Z^{A}\ff{\p}{\p \theta^A}\int_{0}^{1} \ff{dt}{t} f(tZ;t \theta)\,,\\
&hf(Z;\theta) = f(0;0)\,.
\end{align}
The nilpotency of $\Delta$ can be check straightforwardly.
However, in this case \eqref{d*D} does not hold due to the terms
 $\tfrac{1}{2}R^{AB}\tfrac{\partial^{2}}{\partial Y^{A}\partial\theta^{B}}$
 and $ie^{AB}\tfrac{\partial^{2}}{\partial Z^{A}\partial Y^{B}}$
  in \eqref{D}, that do not anticommute with $\dr^{*}$.

To bypass this obstacle we perform a similarity transformation on
the space of functions in question
\be\label{transf}
\
f\left(Z;Y;\theta\right)\longrightarrow\widetilde{f}\left(Z;Y;\theta\right)=\exp\left\{
-e^{AB}\dfrac{\partial^{2}}{\partial
Y^{A}\partial\theta^{B}}\right\} f\left(Z;Y;\theta\right)\,.
\ee
In terms of wavy functions \eqref{mainp} turns to
\be\label{mod}
\ D^{L}\widetilde{f}-e^{AB}Y_{A}\dfrac{\partial}{\partial
Y^{B}}\widetilde{f}-2i\dr_{Z}\widetilde{f}+R^{AB}Z_{A}\dfrac{\partial}{\partial\theta^{B}}\widetilde{f}=\widetilde{J}\,.
\ee
Now though containing $R$-term, the operator $\D$ in \eqref{mod}
results in \eqref{d*D} being satisfied, and the general solution
is provided by \eqref{gen_sol}. Performing inverse transformation,
one finds
\be
\ f=\exp\left\{ e^{AB}\dfrac{\partial^{2}}{\partial
Y^{A}\partial\theta^{B}}\right\} \dr^{*}\exp\left\{
-e^{AB}\dfrac{\partial^{2}}{\partial
Y^{A}\partial\theta^{B}}\right\} J+g+\Delta_{ad}\epsilon\,,
\ee
where $g(Y)$ solves
\be
\ D^{L}g+e^{AB}Y_{A}\dfrac{\partial}{\partial
Y^{B}}g=h\left(\exp\left\{ -e^{AB}\dfrac{\partial^{2}}{\partial
Y^{A}\partial\theta^{B}}\right\} J\right)\,.
\ee
Substituting \eqref{d*} and simplifying, one recovers \eqref{ad}
and \eqref{Had}. Applying resolution of identity \eqref{res_d} to
some $\widetilde{f}$ one finds that in terms of corresponding $f$
it amounts to
\be
\{\Delta_{ad}, \Delta^{-1}_{ad}\}+\mathcal{H}_{ad}=Id\,.
\ee

Formula \eqref{ad} enables us to complete vacuum solution
\eqref{vac}. Indeed at the lowest order we have
\be
\Delta_{ad}\W_0^{3}=ig\gga*\bar{\gga}
\ee
and since $\mathcal{H}_{ad}(\gga*\bar{\gga})=0$, from \eqref{ad}
one finds
\begin{align}\label{vacW3}
\W_{0}^{3}&=-2gZ^{A}\ff{\p}{\p \theta^A}\int_{0}^{1}dt t^3
e^{itZ_{A}Y^{A}+i(1-t)\ff12Z_{A}e^{AB}\ff{\p}{\p\theta^B}}\gd^{4}(\theta)=\notag\\
&=-2gZ^A\int_{0}^{1}dt t^3 e^{itZ_A
Y^A}\gd_{A}'\left(\theta_B+i(1-t)\ff 12 e_{B}{}^{C}Z_C\right)k\bar
k\,.
\end{align}
Formally, this expression for the vacuum 3-form coincides with
that of \cite{DMV} with Lorentz connection $\go^L_{AB}$ set to
zero. But these two expressions represent partial solutions for
two different systems of equations, related by Stueckelberg
transformation \eqref{St1}, \eqref{St2}. Thus, in general, one
could expect that, for instance, the curvature $R^{AB}$ will
appear in \eqref{vacW3}, as it enters \eqref{St2}. However, this
does not happen because, as we have shown, $R$-dependent terms do
not contribute to homotopy operators. So terms with Lorentz
curvature may appear in the vacuum 3-form only via
$\Delta_{ad}$-exact additions to \eqref{vacW3}. It can be shown
however that \eqref{vacW3} is literally reproduced from that of
\cite{DMV} by Stueckelberg transformation \eqref{St2}.

Analogously in the twisted-adjoint case the perturbative operators
repeat those from \cite{DMV} with $\go^L_{AB}=0$ leading to a
significant simplification of the operator form as we now show.

\subsection{Twisted-adjoint case}
In this case $f\sim f_1 k+ f_2 \bar k$. Separating again the
dependence on $k$ and $\bar k$ and abusing notation renaming
$f_{1,2}\to f$ one finds that commutator $[W_0,f]$ turns into
anticommutator
\be
\ff i2e^{\al\dal}\{y_{\al}\bar{y}_{\dal}, f\}=\ff i2e^{AB}\left(
Y_AY_B-\ff{\p^2}{\p Y^A\p Y^B}-2iY_A\ff{\p}{\p Z^B}-\ff{\p^2}{\p
Z^A\p Z^B}\right)
\ee
so that
\begin{align}
\D_{tw}=&\dr+\go_{L}^{AB}\left( Y_{A}\ff{\p}{\p
Y^B}+Z_{A}\ff{\p}{\p Z^B}+\theta_{A}\ff{\p}{\p
\theta^B}\right)-\ff{i}{2}R^{AB}\left(
Z_{A}\ff{\p}{\p\theta^B}+i\ff{\p}{\p Y^A}\ff{\p}{\p
\theta^B}\right)+\\
+&\ff i2e^{AB}\left( Y_AY_B-\ff{\p^2}{\p Y^A\p
Y^B}-2iY_A\ff{\p}{\p Z^B}-\ff{\p^2}{\p Z^A\p Z^B}\right)\,.
\end{align}
Once again, the contribution that does not anticommute with
$\dr^{*}_{Z}$ is
\be
\D_{tw}\sim e^{AB}\left( Y_{A}\ff{\p}{\p Z^B}-\ff i2\ff{\p^2}{\p
Z^A\p Z^B}\right)+
\dfrac{1}{2}R^{AB}\dfrac{\partial^{2}}{\partial\theta^{A}\partial Y^{B}}\,,
\ee
and can be compensated by the following transformation
\be\label{transf_tw}
f\left(Z;Y;\theta\right)\longrightarrow\widetilde{f}\left(Z;Y;\theta\right)=\exp\left\{
ie^{AB}Y_{A}\dfrac{\partial}{\partial\theta^{B}}+\dfrac{1}{2}e^{AB}\dfrac
{\partial^{2}}{\partial Z^{A}\partial\theta^{B}}\right\} f\left(Z;Y;\theta\right)\,.
\ee
For wavy functions one has
\be
D^{L}\widetilde{f}+\dfrac{i}{2}e^{AB}\left(Y_{A}Y_{B}-\dfrac{\partial^{2}}
{\partial Y^{A}\partial Y^{B}}\right)\widetilde{f}-2i\dr_{Z}\widetilde{f}-
\dfrac{i}{2}R^{AB}Z_{A}\dfrac{\partial}{\partial\theta^{B}}\widetilde{f}=\widetilde{J}
 \,,
\ee
that satisfies \eqref{d*D} and thus is solved by \eqref{gen_sol}.

Inversion of \eqref{transf_tw} leads to the following operators
\eqref{sol}
\begin{align}
&\Delta^{-1}_{tw}J=-\ff{1}{2i}Z^A\ff{\p}{\p\theta^A}\int_{0}^{1}\ff{dt}{t}
e^{-i\ff{1-t}{2t}e^{AB}Y_{A}\ff{\p}{\p\theta^B}-\ff{1-t^2}{4t^2}e^{AB}\ff{\p^2}{\p
Z^A\p\theta^B}}J(tZ, Y; t\theta)\,,\notag\\
&\mathcal{H}_{tw}J=e^{-\ff i2e^{AB}Y_A\ff{\p}{\p\theta^B}-\ff
14e^{AB}\ff{\p^2}{\p Z^A\p\theta^B}}J(Z,Y;
\theta)\Big|_{Z=\theta=0}\,,\label{Htw}
\end{align}
which provide a resolution of identity
\be
\{\Delta_{tw}, \Delta^{-1}_{tw}\}+\mathcal{H}_{tw}=Id\,,
\ee
determining the perturvative expansion in the twisted-adjoint
sector.

An important comment is as follows. The elaborated
perturbation theory is based on the conventional resolution
$d^{*}_{Z}$, \eqref{conhom} which is known to lead to
non-localities in HS interactions at second order on the equations
of motion first observed in \cite{GY1} and later on shown more systematically
 in \cite{Boulanger:2015ova,Vasiliev:2017cae} This raises a serious question on
the class of admissible functions, which is not a subject of current
investigation. Nevertheless at least from technical stand point
the conventional resolution can be a useful tool in a search for
the aforementioned class of functions. For example in
\cite{Vasiliev:2016xui} and \cite{Gelfond:2017wrh} it was used in
derivation of local HS cubic interactions.

\section{Conclusion}
\label{con} Let us summarize the main findings of this work. It is
shown that HS extended equations of \cite{Vasiliev:1504} having
room for gauge invariant functionals admit local Lorentz symmetry.
This symmetry gets manifest upon field redefinition leading to the
explicitly Lorentz covariant form of HS equations. In obtaining
Lorentz frame we notice that its consistency condition is more
severe that the standard unfolded one ruling out certain HS
systems that otherwise are formally consistent within the unfolded
frame. This fact provides a double check of the central
element 4-form in \eqref{ehs1} chosen in [28] when constructing
generalized HS equations containing invariant functional. Unlike
the 2-form central element of the original HS equations
\eqref{hs1} which can be tested by its consistency with free HS
equations, the central 4-form can not be checked by the
linearization as it belongs to a higher-form sector responsible
for the invariant functional rather than dynamical equations.

The existence of Lorentz frame for system \eqref{ehs1},
\eqref{ehs2} though mandatory due to equivalence principle is not
quite trivial since it imposes an overdetermined set of
constraints being all fulfilled. Particulary, while higher-form
extension of the equations of \cite{more} admits certain freedom
in a top-form central term from mere consistency, the only one of
those is compatible with the obtained Lorentz covariant equations.
A key element of the construction intrinsically responsible for
Lorentz symmetry is the twistor sector (algebraic constraints for
space-time 0-form fields) of the HS extended equations. The
twistor sector of the original HS equations is known to generate
Lorentz symmetry through the deformed oscillators. So does the
extended twistor sector \eqref{tw1}, \eqref{tw5}. Supporting
Lorentz symmetry it represents a higher-dimensional generalization
of the deformed oscillators underling nonlinear HS equations which
deserves special attention in the context of Calogero-like models
and perhaps higher-dimensional HS equations.

A byproduct of the obtained Lorentz frame is a significant
simplification of the HS perturbation theory despite seemingly
involved form of the covariant equations. We have reconsidered the
analysis of \cite{DMV} and found the covariant version of the
perturbative series operators to gain a simpler form within the
new approach. This significantly simplifies calculation of the
invariant functional density 4-form at least to lowest orders.
\section*{Acknowledgements}

The research was supported by the Russian Science Foundation grant
14-42-00047 in association with Lebedev Physical Institute. The work of
N.G.M. was partially supported by the Foundation for
Theoretical Physics Development ``Basis''.


\begin{thebibliography}{20}
\bibitem{Vasiliev:1990pl} M. A. Vasiliev, {\it Phys.Lett.} {\bf B243} (1990) 378�-382.

\bibitem{more} M. A. Vasiliev, {\it Phys.Lett.} {\bf B285} (1992) 225�-234.

\bibitem{Vasiliev:88} M. A. Vasiliev, {\it Phys.Lett.} {\bf B209} (1988) 491-�497.

\bibitem{Vasiliev:89} M. A. Vasiliev, {\it Annals Phys.} {\bf 190} (1989) 59�-106.

\bibitem{Vasiliev:Rev} M. A. Vasiliev, {\it Higher spin gauge theories: Star-product and AdS
space,} [hep-th/9910096].

%\cite{Boulanger:2008up}
\bibitem{Boulanger:2008up}
  N. Boulanger, C. Iazeolla and P. Sundell,
  %``Unfolding Mixed-Symmetry Fields in AdS and the BMV Conjecture: I. General Formalism,''
  {\it JHEP} {\bf 0907} (2009) 013
  %doi:10.1088/1126-6708/2009/07/013
  [arXiv:0812.3615].

\bibitem{Fronsdal} C. Fronsdal, {\it Phys.Rev.} {\bf D18} (1978) 3624; {\bf D20} (1979) 848.

%\cite{Berends:1984rq}
\bibitem{Berends:1984rq}
  F. A. Berends, G. J. H. Burgers and H. van Dam,
  %``On the Theoretical Problems in Constructing Interactions Involving Higher Spin Massless Particles,''
  {\it Nucl.Phys.} {\bf B260} (1985) 295.
 % doi:10.1016/0550-3213(85)90074-4


\bibitem{Manvelyan:2010jr}
  R.Manvelyan, K.Mkrtchyan and W.Ruhl,
  %``General trilinear interaction for arbitrary even higher spin gauge fields,''
  Nucl.\ Phys.\ B {\bf 836} (2010) 204  [arXiv:1003.2877].

\bibitem{Manvelyan:2010je} R. Manvelyan, K. Mkrtchyan and W. Ruehl, {\it Phys.Lett.} {\bf B696} (2011) 410 [arXiv:1009.1054].

%\cite{Joung:2012rv}
\bibitem{Joung:2012rv}
  E. Joung, L. Lopez and M. Taronna,
  %``On the cubic interactions of massive and partially-massless higher spins in (A)dS,''
  {\it JHEP} {\bf 1207} (2012) 041
 % doi:10.1007/JHEP07(2012)041
  [arXiv:1203.6578].
  %%CITATION = doi:10.1007/JHEP07(2012)041;%%
  %53 citations counted in INSPIRE as of 16 Oct 2017


  %\cite{Francia:2016weg}
\bibitem{Francia:2016weg}
  D. Francia, G. L. Monaco and K. Mkrtchyan,
  %``Cubic interactions of Maxwell-like higher spins,''
  {\it JHEP} {\bf 1704} (2017) 068
 % doi:10.1007/JHEP04(2017)068
  [arXiv:1611.00292].

\bibitem{Sleight:2016dba}
  C.~Sleight and M.~Taronna,
  %``Higher Spin Interactions from Conformal Field Theory: The Complete Cubic Couplings,''
  Phys.\ Rev.\ Lett.\  {\bf 116} (2016) no.18,  181602
  [arXiv:1603.00022].

%\cite{Bengtsson:1983pd}
\bibitem{Bengtsson:1983pd}
  A. K. H. Bengtsson, I. Bengtsson and L. Brink,
  %``Cubic Interaction Terms for Arbitrary Spin,''
 {\it  Nucl.Phys.} {\bf B227} (1983) 31.


\bibitem{Bengtsson:1987}
A. K. H. Bengtsson, I. Bengtsson and N. Linden,
{\it Class.Quant.Grav.} {\bf 4} (1987) 1333.


\bibitem{Metsaev:1993}
R. R. Metsaev, {\it Mod.Phys.Lett.} {\bf A8} (1993) 2413�-2426.


\bibitem{Metsaev:2006}
R. R. Metsaev, {\it Nucl.Phys.} {\bf B759} (2006) 147�-201,
[hep-th/0512342].


%\cite{Metsaev:2007rn}
\bibitem{Metsaev:2007rn}
  R. R. Metsaev,
  %``Cubic interaction vertices for fermionic and bosonic arbitrary spin fields,''
{\it Nucl.Phys.} {\bf B859} (2012) 13
 % doi:10.1016/j.nuclphysb.2012.01.022
  [arXiv:0712.3526].
  %%CITATION = doi:10.1016/j.nuclphysb.2012.01.022;%%
  %93 citations counted in INSPIRE as of 16 Oct 2017


\bibitem{Ponomarev:2016}
D. Ponomarev,
{\it JHEP} {\bf 1612} (2016) 117
[arXiv:1611.00361].


\bibitem{Bengtsson:2016}
A. K. H. Bengtsson, {\it JHEP} {\bf 1612} (2016) 134
[arXiv:1607.06659].


\bibitem{Taronna:2017}
M. Taronna, {\it JHEP} {\bf 1705} (2017) 026
[arXiv:1701.05772].


\bibitem{Ponomarev:2017}
D. Ponomarev, E. Skvortsov,
{\it J.Phys.} {\bf A50} (2017) no.9, 095401
[arXiv:1609.04655].

\bibitem{Ponomarev:2017ar}
D. Ponomarev,
\textit{Chiral Higher Spin Theories and Self-Duality},
[arXiv:1710.00270].


\bibitem{Metsaev:1991mt}
  R. R. Metsaev,
  %``Poincare invariant dynamics of massless higher spins: Fourth order analysis on mass shell,''
{\it Mod.Phys.Lett.} {\bf A6} (1991) 359.


%\cite{Vasiliev:1980as}
\bibitem{Vasiliev:1980as}
  M. A. Vasiliev,
  %``'gauge' Form Of Description Of Massless Fields With Arbitrary Spin. (in Russian),''
{\it Yad.Fiz.} {\bf 32} (1980) 855
   [{\it Sov.J.Nucl.Phys.} {\bf 32} (1980) 439].

%\cite{Vasiliev:1986td}
\bibitem{Vasiliev:1986td}
  M. A. Vasiliev,
  %``Free Massless Fields of Arbitrary Spin in the De Sitter Space and Initial Data for a Higher Spin Superalgebra,''
{\it Fortsch.Phys.} {\bf 35} (1987) 741
   [{\it Yad.Fiz.} {\bf 45} (1987) 1784].


%\cite{Fradkin:1987ks}
\bibitem{Fradkin:1987ks}
  E. S. Fradkin and M. A. Vasiliev,
  %``On the Gravitational Interaction of Massless Higher Spin Fields,''
{\it Phys.Lett.} {\bf B189} (1987) 89.

\bibitem{properties}
 M. A. Vasiliev,
{\it Class.Quant.Grav.} {\bf 8} (1991) 1387.


\bibitem{SS:an} E. Sezgin and P. Sundell,
{\it JHEP} {\bf 0207} (2002) 055 [hep-th/0205132].

\bibitem{Vasiliev:1504}
  M. A. Vasiliev,
  %``Invariant Functionals in Higher-Spin Theory,''
  {\it Nucl.Phys.} {\bf B916} (2017) 219
 % doi:10.1016/j.nuclphysb.2017.01.001
  [arXiv:1504.07289].


\bibitem{KP}
 I. R. Klebanov and A. M. Polyakov,
  %``AdS dual of the critical O(N) vector model,''
{\it Phys.Lett.} {\bf B550} (2002) 213
  %doi:10.1016/S0370-2693(02)02980-5
  [hep-th/0210114].


\bibitem{SS} E. Sezgin and P. Sundell,
{\it JHEP} {\bf 0507} (2005) 044
[hep-th/0305040].

\bibitem{LP} R. G. Leigh and A. C. Petkou,
{\it JHEP} {\bf 0306} (2003), 011
[hep-th/0304217].

\bibitem{GY1} S. Giombi and X. Yin,
{\it JHEP} 1009 (2010) 115 [arXiv:0912.3462].

\bibitem{MZh1} J. Maldacena and A. Zhiboedov,
{\it J.Phys.} {\bf A46} (2013) 214011 [arXiv:1112.1016].

\bibitem{GYrev} S. Giombi and X. Yin,
{\it J.Phys.} {\bf A46} (2013) 214003
[arXiv:1208.4036].

%\cite{Koch:2010cy}
\bibitem{Koch:2010cy}
  R. de Mello Koch, A. Jevicki, K. Jin and J. P. Rodrigues,
  {\it Phys.Rev.} {\bf D83} (2011) 025006
  [arXiv:1008.0633].

%\cite{Bae:2016xmv}
\bibitem{Bae:2016xmv}
  J. B. Bae, E. Joung and S. Lal,
  {\it JHEP} {\bf 1612} (2016) 077
  [arXiv:1611.00112].


\bibitem{Ts} A. Tseytlin, {\it Nucl.Phys.} {\bf B877} (2013) 598-631
[arXiv:1309.0785].

\bibitem{GK} S. Giombi and I. R. Klebanov, {\it JHEP} {\bf 1312} (2013) 068 [arXiv:1308.2337].

\bibitem{GKT} S. Giombi, I. R.  Klebanov, and  A. A. Tseytlin,
{\it Phys.Rev.} {\bf D90.2} (2014) 024048 [arXiv:1402.5396].


\bibitem{BS} N. Boulanger and P. Sundell,
{\it J.Phys.} {\bf A44}
(2011) 495402 [arXiv:1102.2219].

\bibitem{DMV} V. E. Didenko, N. G. Misuna and M. A. Vasiliev,
{\it JHEP} {\bf 1607} (2016) 146 [arXiv:1512.04405].


\bibitem{IS} C. Iazeolla and P. Sundell,
{\it JHEP} {\bf 1112}
(2011) 084 [arXiv:1107.1217].

%\cite{Vasiliev:1989qh}
\bibitem{Vasiliev:1989qh}
  M. A. Vasiliev,
  %``Quantization on sphere and high spin superalgebras,''
  {\it JETP Lett.} {\bf 50} (1989) 374
   [{\it Pisma Zh.Eksp.Teor.Fiz.} {\bf 50} (1989) 344].

   %\cite{Vasiliev:1989re}
\bibitem{Vasiliev:1989re}
  M. A. Vasiliev,
  {\it Int.J.Mod.Phys.} {\bf A6} (1991) 1115.


%\cite{Prokushkin:1998bq}
\bibitem{Prokushkin:1998bq}
  S. F. Prokushkin and M. A. Vasiliev,
  {\it Nucl.Phys.} {\bf B545} (1999) 385
  [hep-th/9806236].

%\cite{Gelfond:2018vmi}
\bibitem{Gelfond:2018vmi}
  O.~A.~Gelfond and M.~A.~Vasiliev,
  %``Homotopy Operators and Locality Theorems in Higher-Spin Equations,''
  arXiv:1805.11941 [hep-th].

%\cite{Vasiliev:2016xui}
\bibitem{Vasiliev:2016xui}
  M. A. Vasiliev,
{\it JHEP} {\bf 1710} (2017) 111
[arXiv:1605.02662].

\bibitem{Vasiliev:2017cae} M. A. Vasiliev,
%{\it On the Local Frame in Nonlinear Higher-Spin Equations}, [arXiv:1707.03735].
JHEP {\bf 1801} (2018) 062 [arXiv:1707.03735 [hep-th]].


\bibitem{DGKV} V.E.~Didenko, O.A.~Gelfond, A.V.~Korybut and M.A.~Vasiliev,
arXiv:1807.00001 [hep-th]

%\cite{Boulanger:2015ova}
\bibitem{Boulanger:2015ova}
  N.~Boulanger, P.~Kessel, E.~D.~Skvortsov and M.~Taronna,
  %``Higher spin interactions in four-dimensions: Vasiliev versus Fronsdal,''
  J.\ Phys.\ A {\bf 49} (2016) no.9,  095402
 % doi:10.1088/1751-8113/49/9/095402
  [arXiv:1508.04139 [hep-th]].

 %\cite{Gelfond:2017wrh}
\bibitem{Gelfond:2017wrh}
O. A. Gelfond and M. A. Vasiliev, %{\it Current Interactions from
%the One-Form Sector of Nonlinear Higher-Spin Equations},
Nucl.\ Phys.\ B {\bf 931} (2018) 383 [arXiv:1706.03718].

\end{thebibliography}
\end{document}